\documentclass[a4paper,11pt,fleqn]{article}
\usepackage{amsmath}
\usepackage{amssymb}
\usepackage{theorem}
\usepackage{charter}
\usepackage{mathrsfs} 
\usepackage{euscript}
\usepackage{xcolor}
\usepackage{enumitem}
\usepackage{caption}
\usepackage{multirow}
\usepackage{array}
\usepackage{booktabs} 
\usepackage[usenames,dvipsnames]{pstricks}

\usepackage{algorithm}
\usepackage{algorithmic}
\usepackage{graphicx}	
\DeclareGraphicsExtensions{.pdf,.jpg,.png,.eps}
\usepackage{url}
\newcommand{\email}[1]{\href{mailto:#1}{\nolinkurl{#1}}}
\PassOptionsToPackage{normalem}{ulem}
\usepackage{ulem}

\definecolor{labelkey}{rgb}{0,0.08,0.45}
\definecolor{refkey}{rgb}{0,0.6,0.0}
\definecolor{Brown}{rgb}{0.45,0.0,0.05}
\definecolor{dgreen}{rgb}{0.00,0.49,0.00}
\definecolor{dblue}{rgb}{0,0.08,0.75}
\RequirePackage[colorlinks,hyperindex]{hyperref}
\hypersetup{linktocpage=true,citecolor=dblue,linkcolor=dgreen}
\newtheorem{theorem}{Theorem}[section]

\theoremstyle{plain}{\theorembodyfont{\rmfamily}%
}
\theoremstyle{plain}{\theorembodyfont{\rmfamily}%
}
\theoremstyle{plain}{\theorembodyfont{\rmfamily}%
}
\theoremstyle{plain}{\theorembodyfont{\rmfamily}%
}
\theoremstyle{plain}{\theorembodyfont{\rmfamily}%
}
\theoremstyle{plain}{\theorembodyfont{\rmfamily}%
}
\theoremstyle{plain}{\theorembodyfont{\rmfamily}%
}

\numberwithin{equation}{section}

\input{alphabet}

\newcommand{\RR}{\ensuremath{\mathbb{R}}}
\newcommand{\NN}{\ensuremath{\mathbb N}}

\newcommand{\HH}{\ensuremath{{\mathcal H}}}

\newcommand{\minimize}[2]{\ensuremath{\underset{\substack{{#1}}}
{\mathrm{minimize}}\;\;#2 }}

\newcommand{\Argmind}[2]{\ensuremath{\underset{\substack{{#1}}}%
{\mathrm{Argmin}}\;\;#2 }}

\renewcommand{\le}{\ensuremath{\leqslant}}
\renewcommand{\ge}{\ensuremath{\geqslant}}

\newcommand{\proj}{\operatorname{proj}}
\newcommand{\Card}[1]{\ensuremath{\sharp \, #1}}

\newcommand{\scal}[2]{{\left\langle{{#1}\mid{#2}}\right\rangle}}

\title{\sffamily A Primal-Dual Data-Driven Method for \\
Computational Optical Imaging with a Photonic Lantern}

\author{
Carlos Santos Garcia$^\dagger$, Mathilde Larchev\^eque$^{\dagger\diamond}$, Solal O'Sullivan$^{\dagger\diamond}$, Martin Van Waerebeke$^{\dagger\diamond}$, \\
Robert R. Thomson$^\ddagger$, 
Audrey Repetti$^{\ddagger\star}$, and 
Jean-Christophe Pesquet$^\dagger$ 
\footnote{This work is supported in part by the ANR Chair in AI BRIDGEABLE; the Royal Society of Edinburgh; EPSRC grant EP/X028860/1; the CVN, INRIA/OPIS and CentraleSup\'elec.}
\\[5mm]
\small
\small $^\dagger$ CentraleSup\'elec, Unversit\'e Paris-Saclay, Gif sur Yvette, France\\
\small $^\ddagger$ School of Engineering and Physical Sciences, Heriot-Watt University, Edinburgh, UK\\
\small $^\star$ School of Mathematical and Computer Sciences, Heriot-Watt University, Edinburgh, UK\\
\small $^\diamond$ Equal contributions \\
\small \texttt{r.r.thomson@hw.ac.uk}, \texttt{a.repetti@hw.ac.uk}, \texttt{jean-christophe@pesquet.eu}
}

\date{}

\begin{document}
\maketitle

\vskip 8mm

\begin{abstract}
Optical fibres aim to image \textit{in-vivo} biological processes. In this context, high spatial resolution and stability to fibre movements are key to enable decision-making processes (e.g., for microendoscopy). 
Recently, a single-pixel imaging technique based on a multicore fibre photonic lantern has been designed, named computational optical imaging using a lantern (COIL). A proximal algorithm based on a sparsity prior, dubbed SARA-COIL, has been further proposed to solve the associated inverse problem, to enable image reconstructions for high resolution COIL microendoscopy. 
In this work, we develop a data-driven approach for COIL. We replace the sparsity prior in the proximal algorithm by a learned denoiser, leading to a plug-and-play (PnP) algorithm. 
The resulting PnP method, based on a proximal primal-dual algorithm, enables to solve the Morozov formulation of the inverse problem.
We use recent results in learning theory to train a network with desirable Lipschitz properties, and we show that the resulting primal-dual PnP algorithm converges to a solution to a monotone inclusion problem.
Our simulations highlight that the proposed data-driven approach improves the reconstruction quality over variational SARA-COIL method on both simulated and real data.
\end{abstract}


\section{Introduction}

\begin{figure}[!t]
    \centering
    \includegraphics[width=0.6\columnwidth]{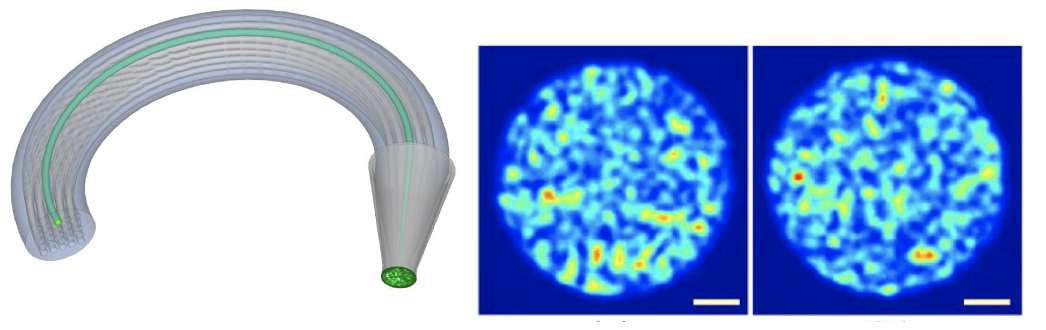}

    \vspace*{-0.3cm}
    
    \caption{\label{fig:MCF}
    (left) Schematic representation of the MCF-PL. (right) Two examples of multimode pattern images obtained through the
MCF.}
    \vspace*{-0.4cm}
\end{figure}

Optical fibres are used for imaging in-vivo biological processes, in particular for microendoscopy. 
To enable decision-making processes for in-vivo observations, the fibre must be stable to movements (e.g., bending), and enable to produce accurate imaging (with high spatial resolution).
On the one hand, standard single-fibre coherent fibre bundles can provide resolutions of a few microns, and can facilitate observation of disease processes at the cellular level when combined with fluorescent contrast agents \cite{Wood2018, Akram2018}. Nevertheless, they are limited either in resolution or in stability. 
Instead, multimode fibres can be used, that can potentially deliver an order of magnitude higher spatial resolution, but they often encounter calibration issues when bending the fibre. 
On the other hand, multicore fibres coupled with a photonic lantern (MCF-PL) have recently been developed, to enable high resolution imaging and robustness to fibre movement \cite{Birks2015, Choudhury2020} (see left part of Fig.~\ref{fig:MCF}).
Distinct multimode light patterns are projected at the output of the lantern by individually exciting the single-mode MCF cores. 
Examples of patterns are given in Fig.~\ref{fig:MCF} (right). 
Imaging through a MCF-PL leads to an ill-posed linear inverse problem, where the objective is to estimate an original unknown image from the photons detected by the single-pixel detector (see \textit{Background} Section).

In \cite{Choudhury2020}, an iterative variational method (SARA-COIL) is proposed to estimate the original image from the MCF-PL measurements. It is based on the primal-dual Condat-V\~u iterations \cite{condat2013primal, vu2013splitting}, for solving a sequence of constrained problems in a redundant wavelet domain. The authors show that it enables accurate reconstruction on simulated data. However, a drop on the reconstruction quality was observed on real data. Further, SARA-COIL is based on a reweighting-$\ell_1$ scheme approximating a log-sum prior \cite{candes2006compressive}, coupled with multiple wavelet transforms, leading to an overall computationally expensive approach with weak convergence guarantees.

We aim to improve the reconstruction quality, yet ensuring the reliability of the solution. Motivated by~the good performance of hybrid optimization method involving neural network (NNs) in computational imaging \cite{adler2018learned, teodoro2019, ahmad2020plug, zhang2020plug}, we develop a new plug-and-play (PnP) algorithm based on primal-dual iterations. The proposed approach solves a constrained problem, using a learned denoising NN to replace the sparsity prior. 
We train the NN to hold desirable Lipschitz properties ensuring the stability of the provided solution \cite{pesquet2021learning}. In addition, we show that the limit point of the proposed algorithm is a zero of a monotone operator.
The proposed approach outperforms SARA-COIL on both simulated and real data.

\paragraph{\textbf{Main contributions.}}
The first main contribution of this work is to demonstrate the efficacy of a data-driven PnP algorithm for the recently developed COIL modality.
The second contribution is to develop a PnP strategy for solving the Morozov formulation of the problem (i.e., by addressing the degradation model through a constraint), which  requires a primal-dual formulation. Compared with Tikhonov formulation, the known advantage of a Morozov approach is to make the choice of hyperparameters easier. 
The third contribution is to provide theoretical convergence guarantees for the provided PnP primal-dual formulation. These guarantees are grounded on the previous recent work of some of the authors~\cite{pesquet2021learning}, which allows to build the denoiser as the learnt resolvent of a maximally monotone operator.
Such a resolvent is a versatile tool which makes the class of considered denoisers quite flexible, while leading to a sound interpretation of the recovered image as a solution to a monotone inclusion problem.
Although many interesting PnP approaches have been developed during the last few years (see e.g., \cite{kamilov2023plug} and references therein), we believe that the two last contributions are pretty original in the field.

We show on simulated and real data that the proposed PnP algorithm outperforms the state-of-the-art variational reweighted-$\ell_1$ approach.

\paragraph{\textbf{Outline.}}
The remainder of the article is organized as follows.
In the \textit{Background} section we describe the formal COIL inverse problem, and provide some background on PnP in the context of monotone inclusion problems.
The proposed approach is presented in Section \textit{Proposed Primal-Dual PnP algorithm}. In Section \textit{Experimental results} we evaluate the performance of the proposed PnP approach on simulated and real data.

\section{Background}
\label{Sec:back}

\subsection{\textbf{Multicore fibre with photonic lantern}}
\label{Ssec:back:COIL}

Light patterns generated by the PL are projected onto an object (e.g., tissue) and light returned from the object (e.g., fluorescence) is detected by a single-pixel detector. 
We consider $M$ multimode light patterns. The $m$-th pattern (for $m\in \{1, \ldots, M\}$) produces a scalar measurement $y_m \in \RR$ corresponding to the sum of the pixelwise multiplication of the pattern and the image of the object of interest. 
Formally, the observations $y \in \eR^M$ are obtained as
\begin{equation} \label{pb:inv}
    y = \Phi \overline{x} + w,
\end{equation}
where $\overline{x} \in \eR^N$ is the unknown image (reshaped to a column vector), 
$ \Phi \in \eR^{M \times N}$ is the linear measurement operator, and $w \in \eR^M$ is a realization of a random perturbation. 
Each row of $\Phi$ contains one pattern (as per Fig.~\ref{fig:MCF}), reshaped to a row vector. The distribution of the noise is not known exactly, but assumed to have a bounded energy, i.e., $\| w \|_2 \le \varepsilon$ for $\varepsilon>0$.

\subsection{\textbf{SARA-COIL methodology}}
\label{Ssec:back:SARACOIL}

The SARA-COIL method \cite{Choudhury2020} defines the estimate as
\begin{equation} \label{pb:min}
    \widehat{x} \in \Argmind{x \in [0,+\infty)^N} f( \Psi x) 
    \text{ subject to } \| \Phi x - y \|_2 \le \varepsilon,
\end{equation}
where $f\circ \Psi$ is a regularization function used to promote sparsity in a transformed domain. 
In~\eqref{pb:min}, $\Psi \in \eR^{S \times N}$ is the concatenation of the first eight Daubechies wavelets transforms and the Dirac basis, so $S=9N$ is the dimension of the transform sparsifying domain, 
and $f \colon \eR^S \to (-\infty, +\infty]$ is a log-sum penalization function \cite{candes2006compressive, candes2007sparsity, carrillo2013SPL} given by
\begin{equation*}
    (\forall v = (v^{(s)})_{1 \le s \le S} \in \RR^S) \quad
    f(v) = \sum_{s=1}^S \log(|v^{(s)}| + \alpha),
\end{equation*}
with $\alpha>0$.
The authors in~\cite{Choudhury2020} propose to solve~\eqref{pb:min} using a reweighting-$\ell_1$ approach \cite{candes2006compressive}, combined with a primal-dual 
algorithm \cite{condat2013primal, vu2013splitting}. Formally, the primal-dual algorithm is used to sequentially solve a collection of problems of the form
\begin{equation} \label{pb:min-l1}
    \widehat{x} \in \Argmind{x \in [0,+\infty)^N} \| \Delta \Psi x \|_1 
    \text{ subject to } \| \Phi x - y \|_2 \le \varepsilon,
\end{equation}
where $\Delta \in \eR^{S \times S}$ is a weight diagonal matrix, whose diagonal elements are chosen according to the current estimate of $\overline{x}$ (see, e.g., \cite{carrillo2013SPL, onose2016scalable} for more details).
SARA-COIL showed a good performance on simulated data, but with a significant drop on real data. 
In addition, this method suffers from the high computational cost mentioned in the introduction.
Finally, although recent works provided theoretical guarantees for some reweighted-$\ell_1$ approaches solving log-sum problems \cite{Geiping2018, Ochs2015siam, Ochs2018JOTA, Repetti2019}, SARA-COIL does not satisfy the necessary conditions to ensure its convergence to a solution to~\eqref{pb:min}.

In this work, we propose a different approach for solving the COIL inverse problem~\eqref{pb:inv}, combining optimization and deep learning, that does not necessitate to use a reweighting procedure.

\subsection{\textbf{Learning MMOs}}

Recently, optimization-based approaches 
have been made more powerful by coupling them with NN models. In particular, unfolded and PnP methods are highly efficient for solving inverse imaging problems \cite{adler2018learned, ahmad2020plug, bertocchi2020deep, jiu2021deep, ren2019simultaneous, teodoro2019, zhang2020plug}. In this work we focus PnP methods. In a nutshell, they consist of replacing some steps in an optimization algorithm by a NN. For image recovery, denoising NNs are often used to replace proximity operators related to the regularization (e.g., replacing the soft-thresholding operator associated with $\ell_1$ regularization).

Multiple works studied the PnP methods theoretical guarantees \cite{cohen2021regularization, Hertrich2021, hurault2022proximal, laumont2022maximum, pesquet2021learning, terris2020building, xu2020provable}. In general, to ensure convergence of the generated iterate sequence, the NN must be firmly non-expansive (FNE). An operator $J \colon \RR^N \to \RR^N$ is FNE \cite{bauschke2017convex} if, for every $(x, y) \in (\RR^N)^2$, $\| J(x) - J(y) \|_2^2 \le \scal{x-y}{J(x) - J(y)}$. 
A few works further provide theoretical characterization of the limit point \cite{cohen2021regularization, Hertrich2021, pesquet2021learning, terris2020building}.

In \cite{pesquet2021learning}, the authors propose to use maximally monotone operator (MMO) theory to design PnP algorithms for solving monotone inclusion problems. 
The objective is to
\begin{equation}    \label{pb:mmo}
    \text{find } \widehat{x} \in \RR^N \text{ such that } 0 \in A(\widehat{x}) + B(\widehat{x}),
\end{equation}
where $A$ and $B$ are MMOs, i.e., for every $(x_1,\! u_1) \!\!\!\!\in \!\!(\RR^N)^2$, $u_1 \!\!\! \in \!\!\! A(x_1)$ if and only if for every $x_2\!\! \in \!\!\!\RR^N$ and $u_2 \!\! \in \!\! A(x_2)$, $\scal{x_1 - x_2}{u_1 - u_2} \! \! \ge  \!0$. 
Such problems can be solved by algorithms grounded on MMO theory, including forward-backward (FB), Douglas–Rachford (DR), primal-dual approaches \cite{bauschke2017convex, combettes2011proximal, combettes2021fixed, condat2013primal, komodakis2015playing, vu2013splitting}.
For these algorithms, $B$ can be handled through its resolvent operator $J_B= (\text{Id} + B)^{-1}$ \cite{bauschke2017convex}. 
Depending on the scheme, operator $A$ is then handled either explicitly (e.g., FB algorithm), through its resolvent operator (e.g., DR algorithm), or using further splitting (i.e., primal-dual methods).


In \cite{pesquet2021learning}, the authors showed that the resolvent $J_B$ of a stationnary MMO $B$ can be approximated by a learned feedforward NN $\widetilde{J}$. They hence paved the way to using algorithms grounded on MMO theory in a PnP fashion, with theoretical guarantees. The authors showed that if $\widetilde{J}$ is FNE, then any sequence $(x_n)_{n\in \mathbb N}$ generated by the resulting PnP algorithm converges to a limit point $\widehat{x} $. They also proposed a FB-PnP algorithm for solving~\eqref{pb:mmo}, when $A$ is the gradient of some convex and Lipschitz-differentiable data-fidelity function. They showed that for this FB-PnP algorithm, $\widehat{x} $ satisfies $0 \in A(\widehat{x}) + \gamma^{-1} \widetilde{B}(\widehat{x})$ for $\widetilde{B} = \widetilde{J}^{-1} - \text{Id}$ and $\gamma\in (0,+\infty)$ being the step-size of the FB-PnP algorithm. 
To the best of our knowledge, this MMO/PnP perspective has not been used yet for solving~\eqref{pb:mmo} considering a non-differentiable data-fidelity, and using other iterative schemes than FB.

\section{Proposed Primal-Dual PnP algorithm}
\label{Sec:proposed}

SARA-COIL solves a collection of problems of the form~\eqref{pb:min-l1}, which is a particular case of~\eqref{pb:mmo} with
\begin{equation}
    A = \Phi^\top N_{\Bc_2(y, \varepsilon)} \Phi, 
    \text{ and }
    B = \partial \| \Delta \Psi \cdot \|_1 + N_{[0,+\infty)^N},
\end{equation}
where $\Bc_2(y, \varepsilon)$ is the $\ell_2$-ball centered in $y$ with radius $\varepsilon$, and $\Phi^\top$ denotes the transpose of matrix $\Phi$.
The normal cone $N_S$ of a subset
$S$ of a Hilbert space $\HH$, equipped with an inner product $\scal{\cdot}{\cdot}$,
is defined
as $N_S\colon x \mapsto \{u \in \HH \mid (\forall y \in S)\, \scal{u}{y-x} \le 0\}$ if
$x\in S$, and $\varnothing$ otherwise.

In this work, we propose to learn a generic operator $B$, i.e., 
\begin{equation}    \label{pb:mon-const}
    \text{find } \widehat{x} \text{ such that } 0 \in \Phi^\top N_{\Bc_2(y, \varepsilon)} \Phi(\widehat{x}) + B(\widehat{x}),
\end{equation}
that aims to better regularize the solution, hence removing the need for reweighting.
An efficient algorithm for solving Problem~\eqref{pb:mon-const} is the primal-dual Condat-V\~u algorithm. The convergence proof of this algorithm is based on MMO theory, and it has the advantage that it does not require operator $\Phi$ to be inverted.

In this work, similarly to \cite{pesquet2021learning}, we propose to characterize $B$ through its resolvent.
Let $\widetilde{J}_\theta$ be an operator
parameterized by a vector $\theta$. If
$\widetilde{J}_\theta = \frac{\text{Id} + Q_\theta}{2}$, where $Q_\theta$ is a $1$-Lipschitz operator,
then $\widetilde{J}_\theta$ is the resolvent
of a MMO $\widetilde{B}$. By modelling $Q_\theta$ as a NN, 
the model parameter vector $\theta$ can thus be learnt to provide an optimal choice for
the MMO regularizer.
The resulting PnP method is given in Algorithm~\ref{algo:PD-mmo-nn}. Then, the following convergence result naturally follows from~\cite{condat2013primal, vu2013splitting} .
\begin{theorem}\label{thm:PD-cvgce}
    Let $(x_k)_{k\in \NN}$ be a sequence generated by Algorithm~\ref{algo:PD-mmo-nn}. Assume that $\tau \sigma \| \Phi \|^2 <1 $, and that $\widetilde{J}_\theta$ is chosen as described above.
    Let $\widetilde{B}$ be the MMO equal to $\widetilde{J}_\theta^{-1} - \operatorname{Id}$. Assume that there exists at least a solution $\widehat{x}$ to the inclusion 

    \vspace{-0.3cm}
    
    \begin{equation} \label{pb:mon-const-nn}
        0 \in \Phi^\top N_{\Bc_2(y, \varepsilon)} \Phi(\widehat{x}) + \tau^{-1} \widetilde{B}(\widehat{x}).
    \end{equation}
    
    \vspace{-0.1cm}
    
    \noindent Then $(x_k)_{k \in \NN}$ converges to a such a solution.
\end{theorem}
Unlike the convergence results of the primal-dual algorithms presented in~\cite{condat2013primal, vu2013splitting}, the primal step-size $\tau$ acts not only on the convergence profile but also on the set of solutions \eqref{pb:mon-const-nn}.
In particular, this theorem shows that $B$ is equal 
to $\widetilde{B}$ up to the multiplicative
factor $1/\tau$. 
A similar behaviour was observed in~\cite{pesquet2021learning} for the FB-PnP algorithm, as discussed in the previous section.

\begin{algorithm}[t]
\caption{PnP primal-dual algorithm for solving~\eqref{pb:mon-const-nn}\label{algo:PD-mmo-nn}}
\begin{algorithmic} 
\STATE 
Let $(x_0, v_0) \in \RR^N \times \RR^M$ and $(\tau, \sigma ) \in (0, +\infty)^2$.
\FOR{$k = 0, 1, \ldots$}
    \STATE $\widetilde{x}_k = x_k - \tau \Phi^\top u_k$
    \STATE $x_{k+1} = \widetilde{J}_\theta \big( \widetilde{x}_k \big) $ 
    \STATE $ \widetilde{u}_k = u_k + \sigma \Phi (2 x_{k+1} - x_k) $
    \STATE $u_{k+1} = \widetilde{u}_k - \sigma \proj_{\Bc_2(y, \varepsilon)} \big( \widetilde{u}_k / \sigma^{-1} \big)$ 
\ENDFOR
\end{algorithmic}
\end{algorithm}

\section{Experimental results}
\label{Sec:results}

\subsection{\textbf{Training}}

Similarly to \cite{pesquet2021learning}, we train $\widetilde{J}_\theta$, on a denoising task, to 
satisfy the desired 1-Lipschitz condition. 
Let $(\overline{x}_i, z_i)_{i\in \mathbb{I}}$ be a set of pairs of ground truth images $(\overline{x}_i)_{i\in \mathbb{I}}$ and associated noisy images $(z_i)_{i\in \mathbb{I}}$. For every $i\in \mathbb{I}$, we have $z_i = \overline{x}_i + \upsilon w_i$, where $w_i \in \RR^N$ is a realization of an additive standard normal random variable, and $\upsilon>0$.

The vector of parameters $\theta$ of the NN is learned so as to
\begin{multline}    \label{pb:loss}
    \minimize{\theta} 
    \dfrac{1}{ \Card{\mathbb{I}} } \sum_{i \in \mathbb{I}} 
    \Big( \| \widetilde{J}_\theta(z_i) - \overline{x}_i \|_2^2 
    + \lambda \max \big\{ \| \nabla Q_\theta(z_i) \|_S^2, 1-\delta \big\} \Big).
\end{multline}
The first term is the standard $\ell^2$ loss for training denoising NNs, while the second term is a Jacobian regularization introduced in \cite{pesquet2021learning} to ensure that $Q_\theta = 2 \widetilde{J}_\theta - \text{Id}$ is $1$-Lipschitz. 
For every $i\in \mathbb{I}$, $\nabla Q_\theta(z_i)$ denotes the Jacobian of $Q_\theta$ computed at $z_i$, and $(\lambda, \delta) \in (0,+\infty)^2$ are regularization parameters. 
In particular, $\lambda$ aims to balance the contribution of $\ell_2$ loss function with the contribution of the Jacobian regularization, and $\delta$ is a parameter chosen to bound the norm away from $1$. In our experiments, we set $\delta=0.05$, and choose manually $\lambda$ to be the smallest parameter ensuring the $1$-Lipschitz condition on $Q_\theta$.
The spectral norm $\| \nabla Q_\theta(z_i) \|_S$ is computed using a power method coupled with back-propagation. 
In \cite{pesquet2021learning}, it was shown that in practice, when $\lambda$ is chosen large enough, then $\| \nabla Q(x) \|<1$ for $x$ belonging to a neighbourhood of the training dataset, which is a sufficient condition to ensure the convergence of Algorithm~\ref{algo:PD-mmo-nn} (see Theorem~\ref{thm:PD-cvgce}).

We train three DnCNN networks \cite{zhang2017dncnn} considering different noise levels $\upsilon \in \{5, 10, 20\}$, on NVIDIA GeForce RTX 2080 Ti GPUs provided by \cite{Fix2022}. We use Adam \cite{adam2014} optimizer on $60\%$ images of the ImageNet \cite{imagenet} validation dataset, converted to grayscale images with values in $[0, 255]$, with batch size $40$, patch size $64 \times 64$, learning rate $ 5\times10^{-5}$, and a scheduler reducing the learning rate by $10\%$ when validation runs without improvement. 
We further use $20\%$ of the ImageNet validation dataset as test set, and the remainder is used for validation during the training.

Training results are summarized in Table~\ref{tab:res-training}, where the $10,000$ images of our test dataset are used to check that $\max_x \| \nabla Q(x) \|_S \le 1$ and evaluate the PSNR values for the denoising task.
For each value of $\upsilon$, we choose the smallest $\lambda$ needed to ensure that the previous inequality holds.

\begin{table}[!t]
    \caption{Training results on denoising problem obtained over the $10,000$ images of the test dataset.}\vspace{-0.2cm}
    \label{tab:res-training}
    \centering
    \begin{tabular}{c c r r}
        $\upsilon$ & $\lambda$ & $\max_x \| \nabla Q(x) \|$ & PSNR (dB)  \\
        \hline
        $5$ & $10^{-3}$ & $0.9963$ & $36.65$ \\
        $10$ & $5\times 10^{3}$ & $0.9952$ & $32.12$ \\
        $20$ & $10^{-2}$ & $ 0.9797 $ & $28.40$ \\
        \hline
    \end{tabular}

\end{table}

\subsection{Experimental setting}
\label{Ssec:sim:setting}

We use the same MCF-PL setting as in \cite{Choudhury2020, Thomson2020_data}. The PL was added at one end of $\sim 3$m of MCF with $121$ single-mode cores in a $11 \times 11$ square
array. 
Each core was individually excited using coherent $514$nm laser light, generating $11^2 =121$ different multimode patterns of light. 
To augment the total measurement number $M$, we also consider a setting where the fiber was rotated $9$ times by $40^\circ$ around the optical axis, creating a total of $121 \times 9 = 1089$ patterns. 
The measurement operator $\Phi$ corresponds to the concatenation of the $M=121$ or $M=1089$ patterns, each of size $N=377 \times 377$.

These patterns are used for both simulated and experimental data. 
For simulated data, measurements are created according to Model~\eqref{pb:inv}, using the $1089$ patterns and a noise level yielding an input SNR of $30$dB. 
For experimental data, measurements were acquired using the fiber and a single-pixel camera, where the object was moved into the beam path, and the magnitude of the light transmitted through the object was recorded by a detector. We highlight the fact that the MCF was intentionally moved and deformed significantly between pattern calibration and imaging experiments, to further highlight the stability of the PL approach (see \cite{Choudhury2020} for more details).
The field of view of all reconstructions using experimentally measured data is $0.9$mm$\times 0.9$mm in the object plane.

\begin{table}[!t]
    \caption{Simulated data: Average results obtained for PnP-COIL over the $50$ simulated data. GPU and CPU time values (in sec.) correspond to the average reconstruction time needed on Intel(R) Core(TM) i9-9940X CPU~@~3.30GHz (RAM 128Gb, 28 cores) with NVIDIA GeForce RTX 2080 Ti (RAM 11Gb).}\vspace{-0.4cm}
    \label{tab:res-simul}
    \centering
    \begin{tabular}{@{}c c c c c@{}}
        $\upsilon$ & PSNR (dB) & SSIM  & {GPU (sec.)}  & CPU (sec.) \\
        \hline
        $5$ & 
        $37.81 (\pm 2.49)$ & $0.698 (\pm 0.012)$ & $13.0 (\pm 1.6)$ &  $70.5 (\pm 8.0)$\\
        $10$ & 
        $37.61 (\pm 2.08) $ & $0.687 (\pm 0.024)$ & $17.0 (\pm 4.3)$ &   {$94.2 (\pm 24.8)$} \\
        $20$ & 
        $36.81 (\pm 2.42)$ & $ 0.672 (\pm 0.022)$ & {$16.4 (\pm 4.9)$} &  {$92.3 (\pm 29.0)$} \\
        \hline
        SARA & $30.72 (\pm 1.38)$ & $0.544 (\pm 0.023)$ & -- & {$98.9 (\pm 11.8)$} \\
        \hline
    \end{tabular}
    \vspace{0.5cm}

\end{table}



\begin{figure}[!t]
    \centering\scriptsize
    \begin{tabular}{ccccc}
        Ground truth $\qquad\;$  &  $\qquad $ SARA-COIL  $ \qquad $ & $\qquad$ PnP -- $\upsilon=5$ $\qquad$  &  $\qquad $  PnP -- $\upsilon=10$ $\qquad$  &  $\qquad $ PnP -- $\upsilon=20$
    \end{tabular}
    
    \includegraphics[width=0.98\columnwidth, trim={0cm 13.3cm 0cm 1.4cm}, clip]{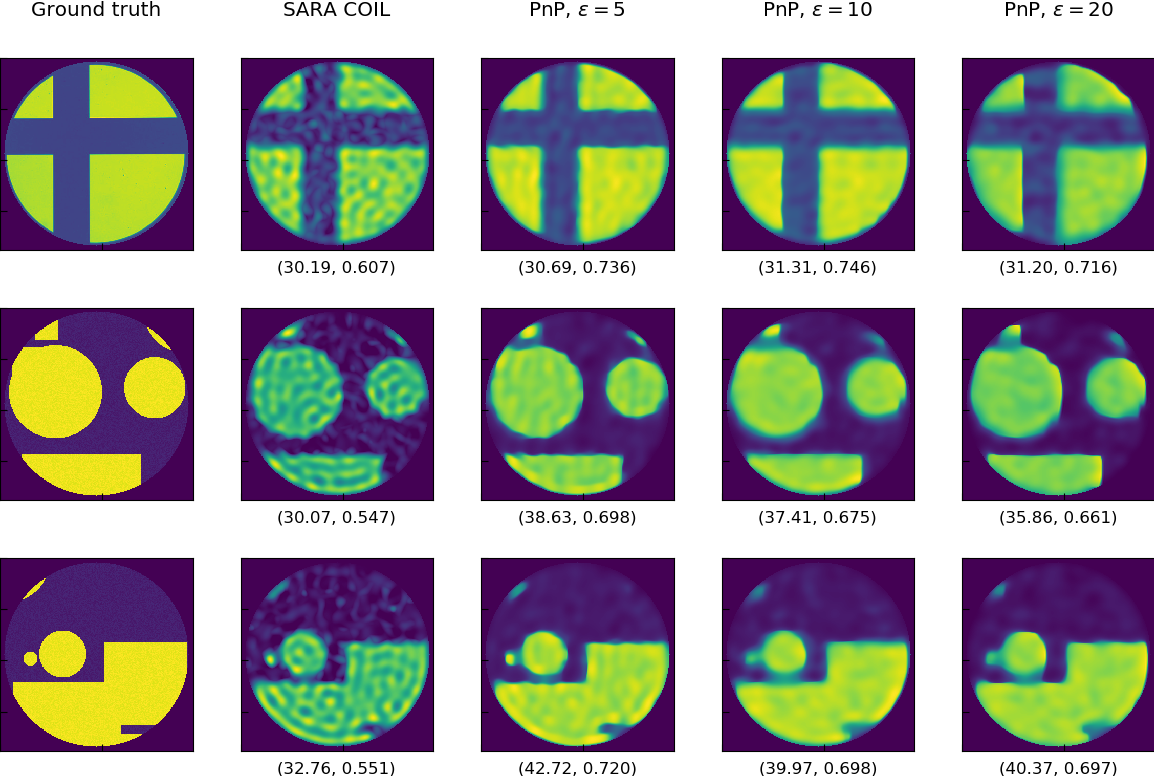}
    
    \begin{tabular}{ccccc}
        $\quad\phantom{aaaaaa}\quad\quad$ $\qquad$  &  $\qquad $ $(30.19, 0.607)$ $\qquad$  &  $\qquad $ $(30.69, 0.736)$ $\qquad$  &  $\qquad $ $(31.31, 0.746)$ $\qquad$  &  $\qquad $ $(31.20, 0.716)$
    \end{tabular}
    
    \includegraphics[width=0.98\columnwidth, trim={0cm 7.0cm 0cm 7.6cm}, clip]{figures/simulated_fig.png}
    
    \begin{tabular}{ccccc}
        $\quad\phantom{aaaaaa}\quad\quad$ $\qquad$  &  $\qquad $ $(30.07, 0.547)$ $\qquad$  &  $\qquad $ $(38.63, 0.698)$ $\qquad$  &  $\qquad $ $(37.41, 0.675)$ $\qquad$  &  $\qquad $ $(35.86, 0.661)$
    \end{tabular}
    
    \includegraphics[width=0.98\columnwidth, trim={0cm 0.6cm 0cm 13.8cm}, clip]{figures/simulated_fig.png}
    
    \begin{tabular}{ccccc}
        $\quad\phantom{aaaaaa}\quad\quad$ $\qquad$  &  $\qquad $ $(32.76, 0.551)$ $\qquad$  &  $\qquad $ $(42.72, 0.720)$ $\qquad$  &  $\qquad $ $(39.97, 0.698)$ $\qquad$  &  $\qquad $ $(40.37, 0.697)$
    \end{tabular}

    \caption{Simulated data: Comparison between ground truth, results from SARA-COIL, and results from proposed PnP-COIL with DnCNNs trained on noise levels $\upsilon \in \{5, 10, 20\}$. Values below images indicate (PSNR, SSIM).}
    \label{fig:simul}

\end{figure}

\subsection{\textbf{Simulated data}}

We validate our primal-dual PnP algorithm on simulated COIL data (dubbed PnP-COIL). 
To this aim, we generate $50$ images with geometric patterns. Examples are shown in Figure~\ref{fig:simul} (left column). 
Average PSNR and SSIM values (with associated standard deviation) are reported in Table~\ref{tab:res-simul}. For these results, we fixed $\varepsilon= 50$ in Algorithm~\ref{algo:PD-mmo-nn}. Quantitative results are very similar for the three trained DnCNNs, showing that PnP-COIL is fairly stable with respect to the training noise level. Visual inspections for three of these images are reported in Figure~\ref{fig:simul}, for images obtained with the three trained DnCNNs and with SARA-COIL \cite{Choudhury2020}. 
We observe that PnP-COIL outperforms SARA-COIL on all examples, although the network has been trained on a very different dataset. Further improvement could certainly be obtained by finetuning the network.
Finally, we observed that PnP-COIL usually requires less iterations than SARA-COIL to reach convergence. 
The average reconstruction time needed for each method is also reported in the last two columns of Table~\ref{tab:res-simul}, when the algorithms make use of GPU or only of CPU, respectively. The CPU reconstruction time is similar for the four methods, although the case $\upsilon=5$ seems to be faster on average. GPU time can only be provided for PnP-COIL\footnote{The SARA-COIL regularization proximity operator is only implemented on CPU, while the DnCNNs plugged in PnP-COIL can run on GPU.}, and yields important accelerations.

\begin{figure}[!t]
    \centering\footnotesize
    \begin{tabular}{ccc} 
    Ground truth & $M=121$ & $M=1089$ \\
    \includegraphics[width=0.2\columnwidth, trim={0cm 7.6cm 29.2cm 1.6cm}, clip]{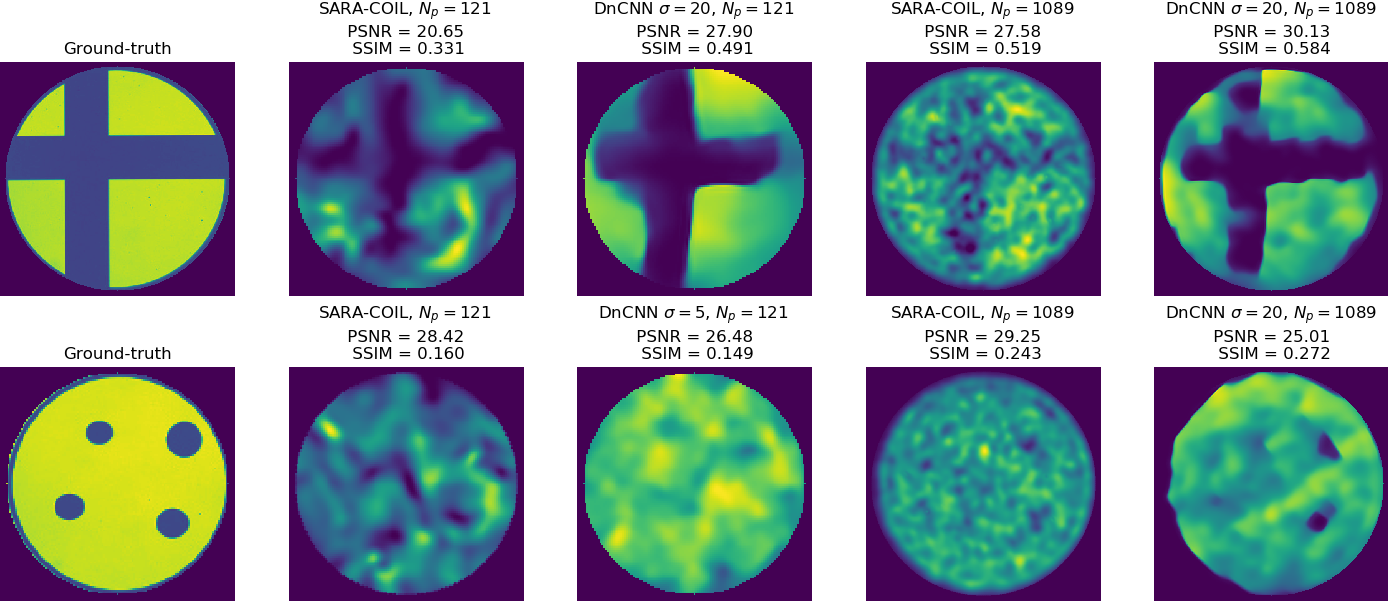}
    &
    \includegraphics[width=0.2\columnwidth, trim={7.3cm 7.6cm 21.9cm 1.6cm}, clip]{figures/final_realdata_results.png}
    &
    \includegraphics[width=0.2\columnwidth, trim={22cm 7.6cm 7.2cm 1.6cm}, clip]{figures/final_realdata_results.png} 
    \end{tabular}
    \begin{tabular}{cc} 
    Ground truth & $M=1089$ \\
    \includegraphics[width=0.2\columnwidth, trim={0cm 0cm 29.2cm 9.2cm}, clip]{figures/final_realdata_results.png}
    &
    \includegraphics[width=0.2\columnwidth, trim={22.0cm 0cm 7.2cm 9.2cm}, clip]{figures/final_realdata_results.png} 
    \end{tabular}
    
    \caption{Experimental data: Ground truth and reconstructions obtained with SARA-COIL considering $M=121$ and $M=1089$ patterns for \textit{cross} image (top) and $M=1089$ patterns for \textit{dots} image (bottom).}
    \label{fig:real_SOTA}

    \vspace*{-0.3cm}
\end{figure}

\begin{figure}[!t]
    \centering\footnotesize
    \begin{tabular}{cccc}
        $ \qquad \varepsilon=2 \qquad $ & $ \qquad \varepsilon=2.5 \qquad $ & $ \qquad \varepsilon=3 \qquad $ & $ \qquad \varepsilon=3.5 \qquad $ \\
        \multicolumn{4}{c}{
        \includegraphics[width=0.98\columnwidth, trim={0.6cm 18.6cm 0 1.8cm}, clip]{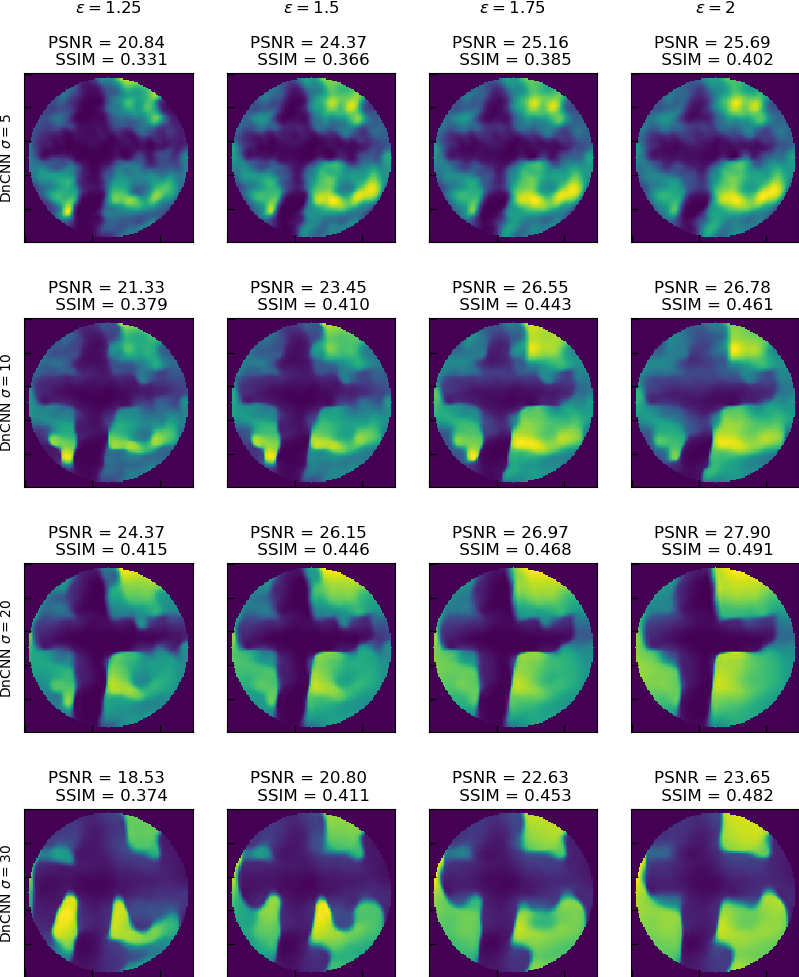}} \\
        $ \qquad \;\; (20.84, 0.331) \; \qquad  $ & $  \qquad \qquad  \; (24.37, 0.366) \;  \qquad $ & $ \qquad  \;\;\; (25.16, 0.470) \;  \qquad $ & $ \qquad  \;\;\;\; (25.69, 0.402) \;  \qquad $ \\
        \multicolumn{4}{c}{
        \includegraphics[width=0.98\columnwidth, trim={0.6cm 12.4cm 0 7.9cm}, clip]{figures/Fullcross121.png}} \\
        $\;\; (21.33, 0.379) \; $ & $ \;\; (23.45, 0.410) \; $ & $ \;\; (26.55, 0.443) \; $ & $ \;\; (26.78, 0.461) \; $ \\
        \multicolumn{4}{c}{
        \includegraphics[width=0.98\columnwidth, trim={0.6cm 6.2cm 0 14.2cm}, clip]{figures/Fullcross121.png}} \\
        $\;\; (24.37, 0.415) \; $ & $ \;\; (26.15, 0.446) \; $ & $ \;\; (26.97, 0.468) \; $ & $ \;\; (27.90, 0.491) \; $ \\
    \end{tabular}
    
    \caption{Experimental data: Reconstructions for \textit{cross} image with $M=121$, obtained with the proposed PnP-COIL, considering different parameters: top to bottom $\upsilon \in \{5, 10, 20\}$, and left to right $\varepsilon \in \{2, 2.5, 3, 3.5\}$.}
    \label{fig:real_cross121}

    \vspace*{-0.3cm}
\end{figure}

\begin{figure}[!t]
    \centering\footnotesize
    \begin{tabular}{cccc}
        $ \qquad \varepsilon=6 \qquad $ & $ \qquad \varepsilon=7 \qquad $ & $ \qquad \varepsilon=8 \qquad $ & $ \qquad \varepsilon=9 \qquad $ \\
        \multicolumn{4}{c}{
        \includegraphics[width=0.98\columnwidth, trim={0.6cm 18.6cm 0 1.8cm}, clip]{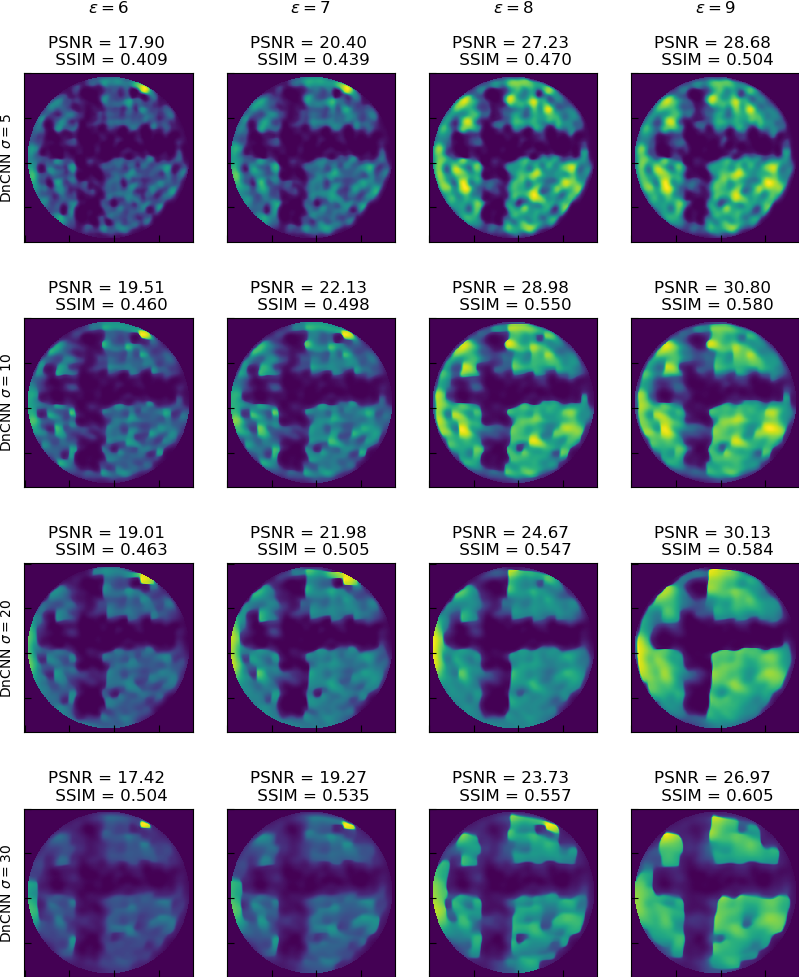}} \\
        $ \qquad \;\; (17.90, 0.409) \;  \qquad $ & $ \qquad \quad  \;\; (20.40, 0.439) \;  \qquad $ & $ \qquad \quad \quad  \; (27.23, 0.470) \;  \qquad $ & $ \qquad  \;\;\;\; (28.68, 0.504) \;  \qquad $ \\
        \multicolumn{4}{c}{
        \includegraphics[width=0.98\columnwidth, trim={0.6cm 12.4cm 0 7.9cm}, clip]{figures/Fullcross1089.png}} \\
        $\;\; (19.51, 0.460) \; $ & $ \;\; (22.13, 0.498) \; $ & $ \;\; (28.98, 0.550) \; $ & $ \;\; (30.80, 0.580) \; $ \\
        \multicolumn{4}{c}{
        \includegraphics[width=0.98\columnwidth, trim={0.6cm 6.2cm 0 14.2cm}, clip]{figures/Fullcross1089.png}} \\
        $\;\; (19.01, 0.463) \; $ & $ \;\; (21.98, 0.505) \; $ & $ \;\; (24.67, 0.547) \; $ & $ \;\; (30.13, 0.584) \; $ \\
    \end{tabular}
    
    \caption{Experimental data: Reconstructions for \textit{cross} image with $M=1089$, obtained with the proposed PnP-COIL, considering different parameters: top to bottom $\upsilon \in \{5, 10, 20\}$, and left to right $\varepsilon \in \{6, 7, 8, 9\}$.}
    \label{fig:real_cross1089}

    \vspace*{-0.3cm}
\end{figure}

\begin{figure}[!t]
    \centering\footnotesize
    \begin{tabular}{cccc}
        $ \qquad \varepsilon=12 \qquad $ & $ \qquad \varepsilon=13 \qquad $ & $ \qquad \varepsilon=14 \qquad $ & $ \qquad \varepsilon=15 \qquad $ \\
        \multicolumn{4}{c}{
        \includegraphics[width=0.98\columnwidth, trim={0.6cm 18.6cm 0 1.8cm}, clip]{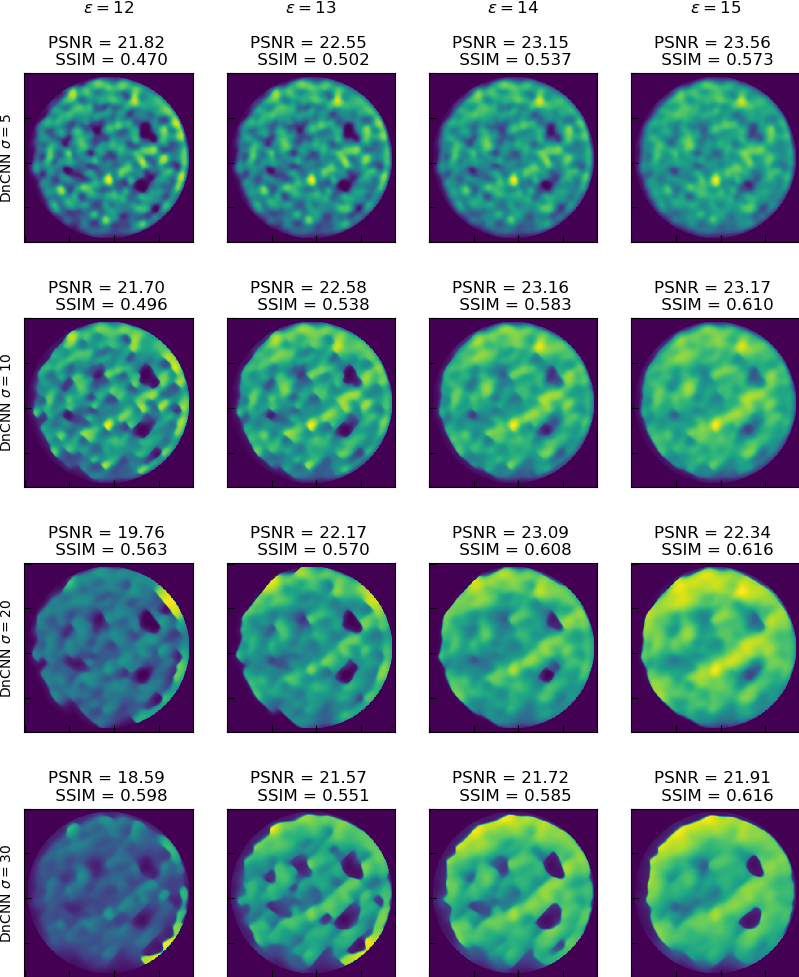}} \\
        $ \qquad \;\; (21.82, 0.470) \;  \qquad $ & $ \qquad \quad \;\; (22.55, 0.502) \;  \qquad $ & $ \qquad \quad \;\;\; (23.15, 0.537) \;  \qquad $ & $ \qquad \quad \;\; (23.56, 0.573) \;  \qquad $ \\
        \multicolumn{4}{c}{
        \includegraphics[width=0.98\columnwidth, trim={0.6cm 12.4cm 0 7.9cm}, clip]{figures/dots1089.png}} \\
        $\;\; (21.70, 0.496) \; $ & $ \;\; (22.58, 0.538) \; $ & $ \;\; (23.16, 0.583) \; $ & $ \;\; (23.17, 0.610) \; $ \\
        \multicolumn{4}{c}{
        \includegraphics[width=0.98\columnwidth, trim={0.6cm 6.2cm 0 14.2cm}, clip]{figures/dots1089.png}} \\
        $\;\; (19.76, 0.563) \; $ & $ \;\; (22.17, 0.570) \; $ & $ \;\; (23.09, 0.608) \; $ & $ \;\; (22.34, 0.616) \; $ \\
    \end{tabular}

    \caption{Experimental data: Reconstructions for \textit{dots} image with $M=1089$, obtained with the proposed PnP-COIL, considering different parameters: top to bottom $\upsilon \in \{5, 10, 20\}$, and left to right $\varepsilon \in \{12, 13, 14, 15\}$.}
    \label{fig:real_dot1089}

    \vspace*{-0.3cm}
\end{figure}

\subsection{\textbf{Experimental data}}

We validate the proposed algorithm on real data, acquired as per Section~\ref{Ssec:sim:setting}, using two images: \textit{cross} and \textit{dots} (Figure~\ref{fig:real_SOTA}-left). The reconstructions obtained with SARA-COIL for \textit{cross} are reported in Figure~\ref{fig:real_SOTA}. Reconstructions using PnP-COIL are given in Figures~\ref{fig:real_cross121}-\ref{fig:real_dot1089}. In each case, we show the results obtained considering the NNs trained on different noise levels $\upsilon$, for different radius of the $\ell_2$-ball $\varepsilon$ in~\eqref{pb:mon-const}. For all cases, the reconstructed images become smoother when both parameters $(\upsilon, \varepsilon)$ increase. For the \textit{cross} example, we observe that the proposed primal-dual PnP method leads to higher accuracy in the reconstruction than SARA-COIL (see Figures~\ref{fig:real_cross121} and \ref{fig:real_cross1089}). For the \textit{dots} example, we see in Figure~\ref{fig:real_dot1089} that the proposed approach enables finding the 4 dots in the image when $M=1089$, while SARA-COIL could barely see one of them.

\section{Conclusion}

In this work, we have introduced a new primal-dual PnP algorithm for solving monotone inclusion problems, in the context of computational optical imaging. The proposed approach enables to handle non-smooth data-fidelity terms involving a linear operator, e.g. an ellipsoidal constraint.
We showed the outperformance of the proposed PnP-COIL approach on simulated and real COIL data with respect to the state-of-the art variational approach.

\section*{Author contributions statement}
{\bf C.S.G.}: Conceptualization, Methodology, Software, Investigation, Visualization.  
{\bf M.L.} Conceptualization, Methodology, Investigation, Visualization.  
{\bf S.O.} Conceptualization, Methodology, Investigation, Visualization. 
{\bf M.V.W.}: Conceptualization, Methodology, Investigation, Visualization. 
{\bf R.R.T.}: Data Curation,  Resources, Validation.
{\bf A.R.}: Conceptualization, Methodology, Software, Data Curation, Writing, Supervision, Project Administration, Funding Acquisition.
{\bf J.-C.P.}: Conceptualization, Methodology, Writing, Supervision, Project Administration, Funding Acquisition.

\section*{Code and data availability}
The data underlying this article are available at \href{DOI:10.17861/a1bebd55-b44f-4b34-82c0-c0fe925762c6}{DOI:10.17861/a1bebd55-b44f-4b34-82c0-c0fe925762c6}.
The codes to reproduce the results of this article (SARA-COIL and PnP-COIL) are available on GitHub \href{https://github.com/carlossantosgarcia/PnP_Lantern}{here}.

\bibliographystyle{IEEEbib}
\small
\bibliography{reference}

\end{document}